\documentclass[aps,twocolumn,showpacs,superscriptaddress,prl]{revtex4}
\usepackage{graphicx,amsmath}
\usepackage[colorlinks,linkcolor=blue,citecolor=blue]{hyperref}
\usepackage{hypcap}

\begin{document}

\title{Spin-orbit Coupled Bose-Einstein Condensates in Spin-dependent
Optical Lattices}
\author{Wei Han}
\affiliation{Beijing National Laboratory for Condensed Matter
Physics, Institute of Physics, Chinese Academy of Sciences, Beijing
100190, China}\affiliation{Institute of Theoretical Physics, Shanxi
University, Taiyuan 030006, China}
\author{Suying Zhang}
\affiliation{Institute of Theoretical Physics, Shanxi University,
Taiyuan 030006, China}
\author{Wu-Ming Liu}
\affiliation{Beijing National Laboratory for Condensed Matter
Physics, Institute of Physics, Chinese Academy of Sciences, Beijing
100190, China}

\begin{abstract}
We investigate the ground-state properties of spin-orbit coupled
Bose-Einstein condensates in spin-dependent optical lattices. The
competition between the spin-orbit coupling strength and the depth
of the optical lattice leads to a rich phase diagram. Without
spin-orbit coupling, the spin-dependent optical lattices separate
the condensates into alternating spin domains with opposite
magnetization directions. With relatively weak spin-orbit coupling,
the spin domain wall is dramatically changed from N\'{e}el wall to
Bloch wall. For sufficiently strong spin-orbit coupling, vortex
chains and antivortex chains are excited in the spin-up and
spin-down domains respectively, corresponding to the formation of a
lattice composed of meron-pairs and antimeron-pairs in the
pseudospin representation. We also discuss how to observe these
phenomena in real experiments.
\end{abstract}

 \maketitle

\textit{Introduction.---}In recent years, the experimental control
on ultracold atomic gases has reached truly unprecedented levels. By
employing two lasers with different frequencies and polarizations
plus a non-uniform vertical magnetic field, experimentalists have
produced spin-orbit (SO) coupling, which couples the internal states
and the orbit motion of the atoms
\cite{Spielman,Jing-Zhang,Zwierlein,Shuai-Chen}. The SO coupled
ultracold atomic gases have attracted great interests of researchers
\cite{Dalibard,Niu,Galitski,Wu,Wu2,Jung-Hoon-Han,Wuming-Liu,Pitaevskii,Baym,Trivedi,Galitski2}.
It has been indicated that the interplay among SO coupling,
interatomic interaction and external potential leads to rich
ground-state phases, such as plane wave, density stripe, fractional
vortex and various vortex lattices
\cite{Wang,Tin-Lun-Ho,You,Santos,Chuanwei-Zhang,Hu,Ramachandhran,Su-Yi,Gou,Ueda,Ruokokoski}.
The SO coupled ultracold atomic gases open a new window for quantum
simulation, and provide opportunities to study SO coupling phenomena
in a highly controllable impurity-free environment.

All the existing studies on SO coupled Bose-Einstein condensates
(BECs) only refer to the case that different internal states of the
atoms are trapped in an identical external potential. However, by
using two counterpropagating lasers with the same frequency but
different polarizations, the experimentalists have been able to
produce spin-dependent optical lattices that allow different
internal states of the atoms experience drastically different
external potentials
\cite{Mandel,McKay,Becker,Soltan-Panahi,Soltan-Panahi2}. The
spin-dependent optical lattices bring more complicated geometry of
condensates and have potential applications in quantum computation
\cite{Mandel}, cooling and thermometry \cite{McKay}, and quantum
simulation \cite{Becker,Soltan-Panahi,Soltan-Panahi2,Hauke}. It is
natural to ask what new structures can be formed due to the
competition between the SO coupling and the spin-dependent optical
lattices.

In this Letter, we investigate the ground-state phase diagram of SO
coupled BECs in spin-dependent optical lattices. In the absence of
SO coupling, the ground state of the system is characterized by the
formation of alternating spin domains. However, such a structure can
be dramatically changed due to the SO coupling effects. Relatively
weak SO coupling basically changes the orientation of the spins in
the domain walls, causing the transformation from N\'{e}el wall to
Bloch wall. Sufficiently strong SO coupling excites meron-pairs and
antimeron-pairs in the spin-up and spin-down domains respectively
and generates a meron-pair lattice. This is essentially different
from the mechanism of generating a meron-pair lattice by bulk
rotation \cite{Kasamatsu2}. Our findings provide a new way to create
and manipulate topological excitations in SO coupled systems.
\begin{figure}[bp]
\capstart\centerline{\includegraphics[width=8.5cm,clip=]{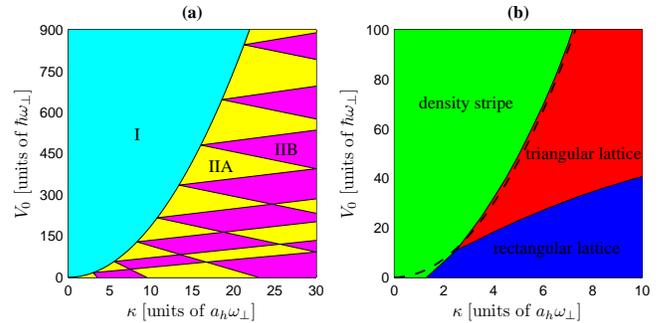}}
\caption{\label{fig1}(Color online) (a) Single-particle phase
diagram spanned by the SO coupling strength $\protect\kappa$ and the
depth $V_{0}$ of the optical lattice. (b) Many-body phase diagram
spanned by $\protect\kappa$ and $V_{0}$ with the effective
interaction parameter $\tilde{g}=6000$. The dashed line in (b)
indicates the phase boundary between phase I and phase II in (a) for
comparison.}
\end{figure}
\begin{figure*}[tbp]
\capstart\includegraphics[width=1.99\columnwidth\vspace{0cm}
\hspace{0cm}]{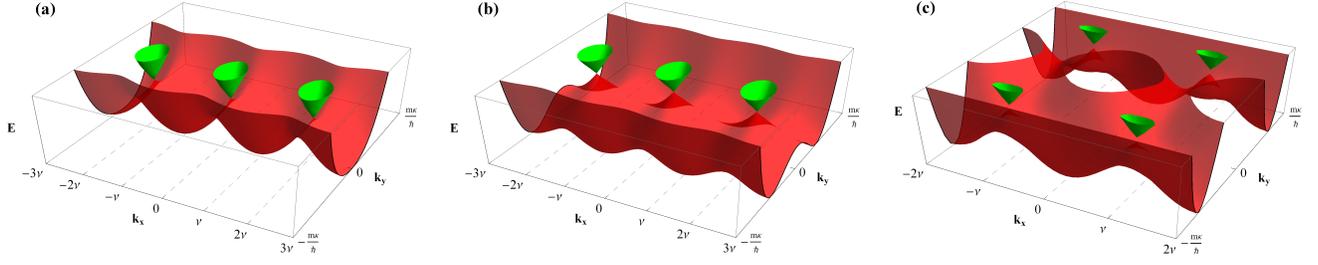}\caption{\label{fig2}(Color online) Band
structures induced by the competition between the SO coupling
strength $\kappa$ and the depth $V_{0}$ of the optical lattice
corresponding to phase I (a), phase IIA (b), and phase IIB (c) of
the single-particle phase diagram in Fig. 1(a). The superposition of
the momenta at the minima of the bands produces a density stripe in
(a), a lattice in (b) and (c).}
\end{figure*}

\textit{Energy band structure.---}We consider SO coupled BECs
confined in the combined potential of a quasi-2D harmonic trap and
1D spin-dependent optical lattices. The Hamiltonian of this system
is given by
\begin{eqnarray}
\mathcal{H} &=&\int d\mathbf{r}\Bigl[\mathbf{\Psi}^{\dag}(-\frac{\hbar ^{2}\boldsymbol{\nabla}^{2}}{2m}+\mathcal{V}_{\text{so}})\mathbf{\Psi} +V_{1}(\mathbf{r}%
)n_{\uparrow}+V_{2}(\mathbf{r})n_{\downarrow}\notag \\
&&+g_{11}n_{\uparrow}^{2}+g_{22}n_{\downarrow}^{2}+g_{12}n_{\uparrow}n_{\downarrow}\Bigr],
\label{Model Hamiltonian}
\end{eqnarray}
where $\mathbf{\Psi} =[\Psi _{\uparrow}(\mathbf{r}),\Psi
_{\downarrow}(\mathbf{r})]^{T}$ denotes the
two-component wave functions and is normalized as $\int d\mathbf{r}%
\mathbf{\Psi}^{\dag}\mathbf{\Psi}=N$ with $N$ the total particle number. $%
n_{\uparrow}=\Psi _{\uparrow}^{\ast }\Psi _{\uparrow}$ and
$n_{\downarrow}=\Psi _{\downarrow}^{\ast }\Psi _{\downarrow}$
describe the particle number density of each component.
$g_{ij}=4\pi\hbar^{2}a_{ij}/m$ ($i,j=1,2$) represent the interatomic
interaction strengths characterized by the $s$-wave scattering
lengths $a_{ij}$ and the atomic mass $m$.

We consider a Rashba SO coupling
$\mathcal{V}_{\text{so}}=-i\hbar\kappa(\sigma_{x}\partial_{x}+\sigma_{y}\partial_{y})$,
where $\sigma_{x,y}$ are the Pauli matrices and $\kappa$ denotes the
SO coupling strength. The combined external potential
$V_{i}(\mathbf{r})=V_{\text{H}}(\mathbf{r})+V_{\text{OL}i}(x)$,
where $V_{\text{H}}=\frac{1}{2}m\omega _{\perp
}^{2}[(x^{2}+y^{2})+\lambda^2z^{2}]$ is the harmonic trapping
potential with
$\lambda=\omega_{z}/\omega_{\perp} \gg 1$, and $V_{\text{OL}1}=V_{0}\sin ^{2}(\nu x)$ and $V_{\text{OL}%
2}=V_{0}\cos ^{2}(\nu x)$ describe the 1D spin-dependent optical
lattice potentials, which are experienced by the two components,
respectively. Approximating the $z$ dependence of the wave functions
by the single-particle ground state in a harmonic potential, one can
obtain the 2D dimensionless effective interaction parameters
$\tilde{g}_{ij}=2\sqrt{2\pi\lambda}Na_{ij}/a_{h}$, with
$a_{h}=\sqrt{\hbar /(m\omega _{\perp })}$ \cite{Kasamatsu}.

The single-particle energy bands are critically important to
understand the ground-state properties of the condensates. Without
considering the harmonic trap, the 2D single-particle wave functions
in the k-space obey the secular equations $\frac{\hbar
^{2}}{2m}[\left( k_{x}+2s\nu \right)
^{2}+k_{y}^{2}]U_{s}-\frac{V_{0}}{4}\left(
U_{s+1}-2U_{s}+U_{s-1}\right) +\hbar\kappa \left(
k_{x}+2s\nu-ik_{y}\right) W_{s}=\varepsilon U_{s}$ and $\frac{\hbar
^{2}}{2m}[\left( k_{x}+2s\nu \right)
^{2}+k_{y}^{2}]W_{s}+\frac{V_{0}}{4}\left(
W_{s+1}+2W_{s}+W_{s-1}\right) +\hbar\kappa \left( k_{x}+2s\nu
+ik_{y}\right) U_{s}=\varepsilon W_{s}$, where $U_{s}=U\left(
k_{x}+2s\nu ,k_{y}\right)$ and $W_{s}=W\left( k_{x}+2s\nu
,k_{y}\right) $ with $s=0,\pm 1,\pm 2,...,\pm \infty $ represent the
wave functions at the point $\left( k_{x}+2s\nu ,k_{y}\right) $.
Typically, we choose $\nu=a_{h}^{-1}$ for our present discussion. By
numerical exact diagonalization, we can solve the secular equations
and obtain the energy band structure.

We find that there exist three different kinds of energy band
structures depending on the competition between the SO coupling
strength $\kappa$ and the depth $V_{0}$ of the optical lattices. Fig.~\ref{fig1}(a) presents the single-particle phase diagram spanned by $%
\kappa $ and $V_{0}$. In phase I, the minima of the energy bands
locate in a set of k points with $k_{y}=0$ and $k_{x} \in
K_{1}=\{\pm \nu,\pm 3\nu,\pm 5\nu,...\}$ [See Fig.~\ref{fig2}(a)].
In phase II, the minima of the energy bands locate in a set of k
points with $k_{y}=\pm \delta $ $\left( 0<\delta\leq
\frac{m\kappa}{\hbar} \right) $, and $k_{x}\in K_{1}$ for phase IIA
($k_{x}\in K_{2}=\{0,\pm 2\nu,\pm 4\nu,\pm 6\nu,...\}$ for phase
IIB) [See Figs.~\ref{fig2}(b) and~\ref{fig2}(c)].
\begin{figure*}[tbp]
\capstart\centerline{\includegraphics[width=16.9cm]{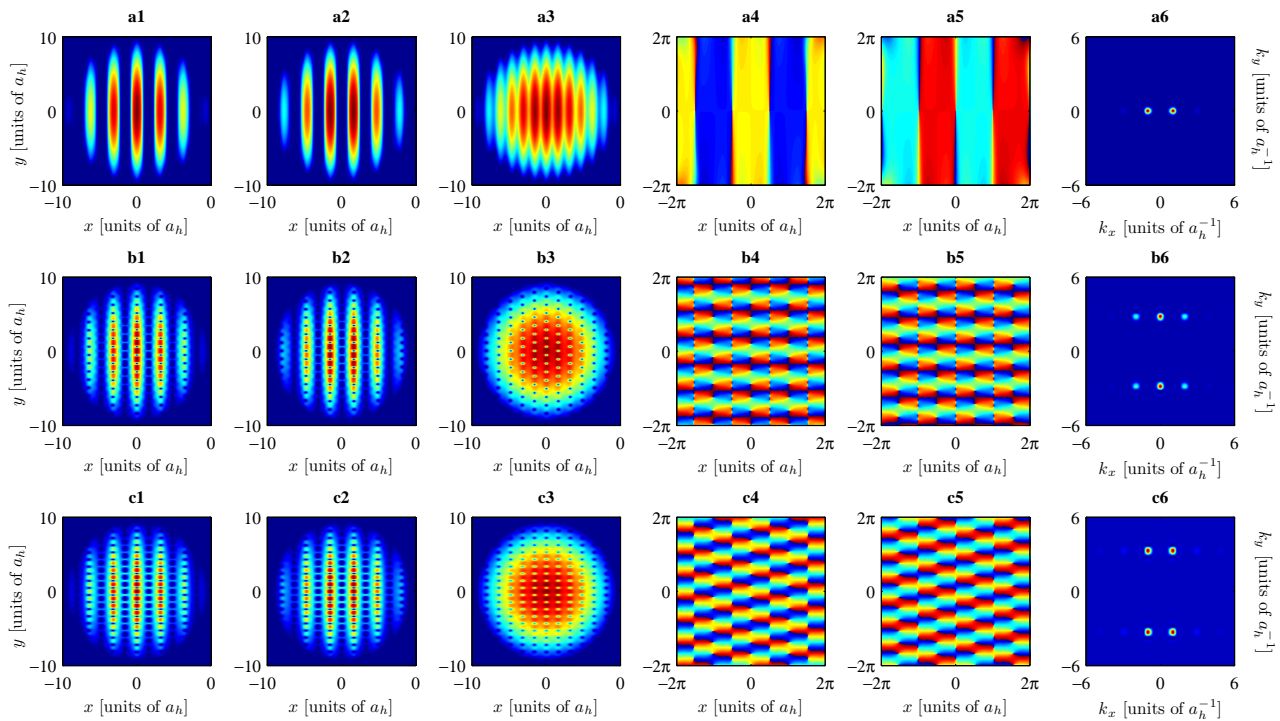}}
\caption{\label{fig3}(Color online) Ground state as a function of
the SO coupling strength $\kappa$ and the depth $V_{0}$ of the
optical lattice with $\kappa=4a_{h}\omega_{\perp}$,
$V_{0}=40\hbar\omega_{\perp}$ (a1-a6),
$\kappa=4a_{h}\omega_{\perp}$, $V_{0}=20\hbar\omega_{\perp}$
(b1-b6), $\kappa=4a_{h}\omega_{\perp}$,
$V_{0}=15\hbar\omega_{\perp}$ (c1-c6). The effective interaction
parameter is fixed at $\tilde{g}=6000$. The spin-up, spin-down, and
total density profiles are shown in (a1, b1, c1), (a2, b2, c2), and
(a3, b3, c3), respectively. The phases of the spin-up and spin-down
wave functions, with values ranging from $-2\pi$ to $2\pi$ (blue to
red), are shown in (a4, b4, c4) and (a5, b5, c5). The momentum
distributions are depict in (a6, b6, c6).}
\end{figure*}

The single-particle ground state in phase I is nondegenerate and can
be expressed as a linear superposition of the plane waves with wave
vectors $(k_{x}\in K_{1},k_{y}=0)$. This yields alternating spin
domains with opposite magnetization directions (density stripe). In
both phase IIA and phase IIB, the single-particle ground state is
double-degenerate. Each degenerate state can be expressed as a
linear superposition of the plane waves with wave vectors $(k_{x}\in
K, k_{y}=\delta)$ or $(k_{x}\in K, k_{y}=-\delta)$, where $K=K_{1}$
for phase IIA and $K=K_{2}$ for phase IIB. For an arbitrary nonzero
superposition of the two degenerate states, lattice will be formed
as the single-particle ground state.

\textit{Phase diagram.---}By using the imaginary time evolution
method, we can solve Eq.~(\ref{Model Hamiltonian}) to obtain the
many-body ground state. Considering that the spin-exchange
integrations are very weak in typical experiments, we just discuss
the case that $\tilde{g}_{ij}=\tilde{g}$. The many-body phase
diagram spanned by $\kappa $ and $V_{0}$ with $\tilde{g}=6000$ is
presented in Fig.~\ref{fig1}(b). We find that the competition
between the SO coupling strength and the depth of the optical
lattice leads to three distinct phases---density stripe, triangular
vortex lattice and rectangular vortex lattice.

In the density stripe phase, the spin-up and spin-down components
are arranged alternately and form alternating spin domains [See
Figs.~\ref{fig3}(a1-a6)]. Comparing the phase diagrams in
Figs.~\ref{fig1}(a) and~\ref{fig1}(b), we find that the interaction
has no significant influence on the phase region of the density
stripe.

In the vortex lattice phases, both the triangular and rectangular
lattices are composed of alternately arranged vortex and antivortex
chains, which are excited in the spin-up and spin-down domains
respectively [See Figs.~\ref{fig3}(b1-b6) and~\ref{fig3}(c1-c6)].
The only difference is that the vortices of the neighboring chains
are staggered for the triangular lattice, but are parallel for the
rectangular lattice [See Figs.~\ref{fig3}(b3) and~\ref{fig3}(c3)].
These two different arrangements of vortices correspond to
odd-parity and even-parity distributions of the particles in $k_{x}$
direction of the k-space. [See Figs.~\ref{fig3}(b6)
and~\ref{fig3}(c6)]. These correspond to the single particle band
structures described in Figs.~\ref{fig2}(c) and (b), where the
minima of the bands also show odd-parity and even-parity
distributions respectively, although their phase regions are not
consistent due to the influence of the interatomic interactions [See
Figs.~\ref{fig1}(a) and~\ref{fig1}(b)]. As discussed above, the
single-particle ground state in phase II is double-degenerate. From
Figs.~\ref{fig3}(b6) and~\ref{fig3}(c6), we can see that the
interaction removes the degeneracy and chooses an equal weighted
linear superposition of the two degenerate states as the many-body
ground state.
\begin{figure}[tbp]
\capstart\includegraphics[width=0.9\columnwidth\vspace{0cm} \hspace{0cm}]{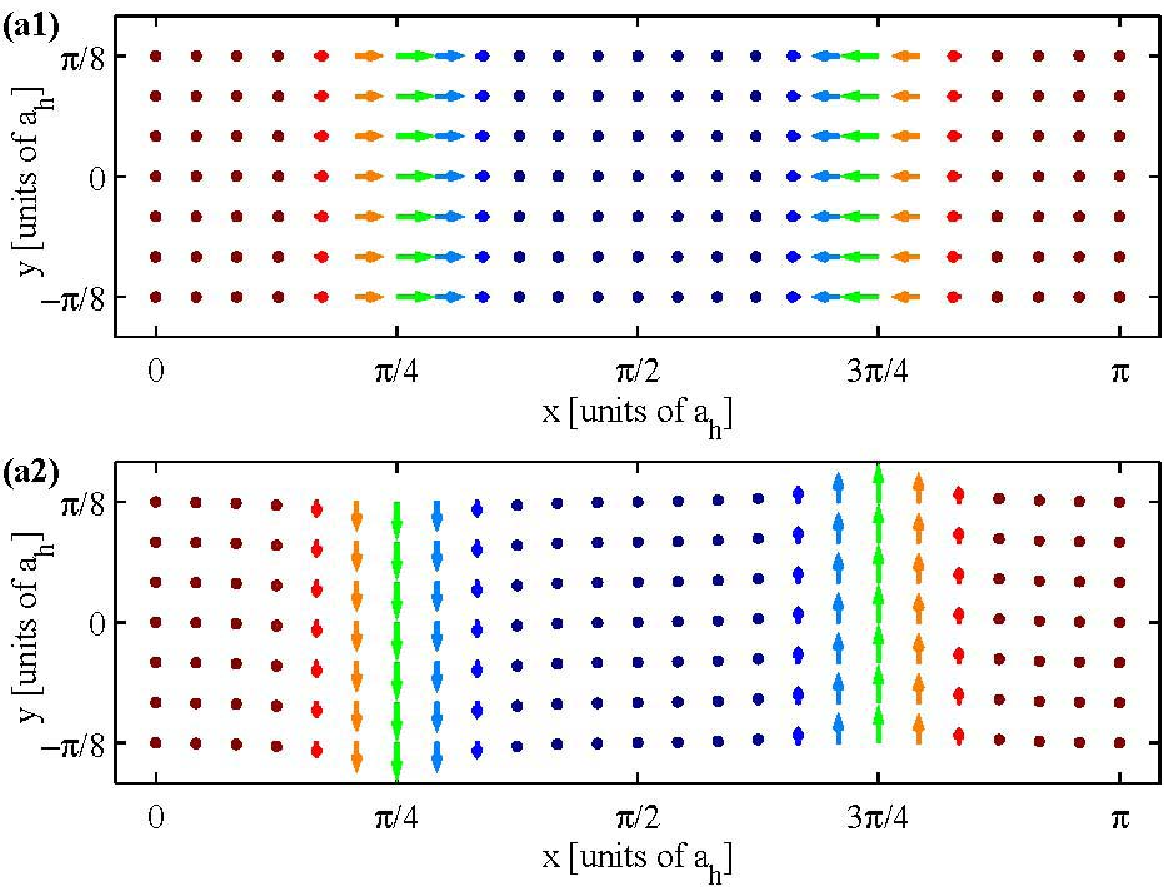}%
\newline
\includegraphics[width=0.9\columnwidth\vspace{0.04cm}
\hspace{0cm}]{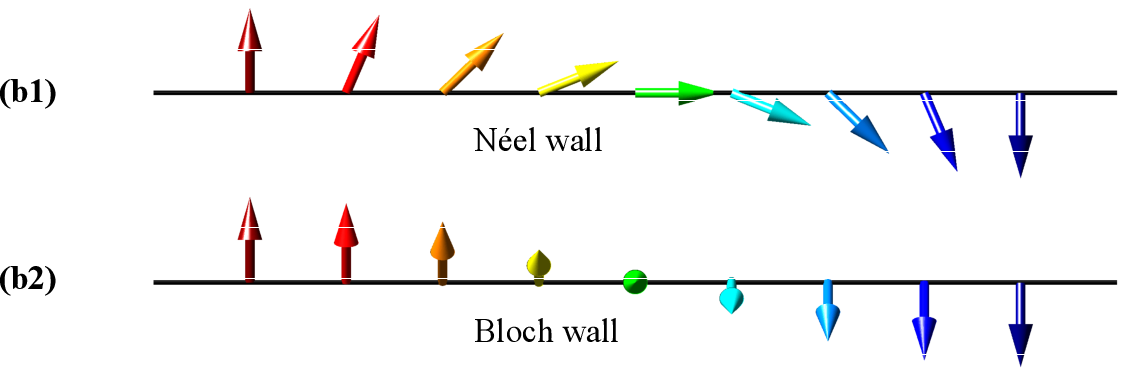}
\includegraphics[width=0.9\columnwidth\vspace{0cm}
\hspace{0cm}]{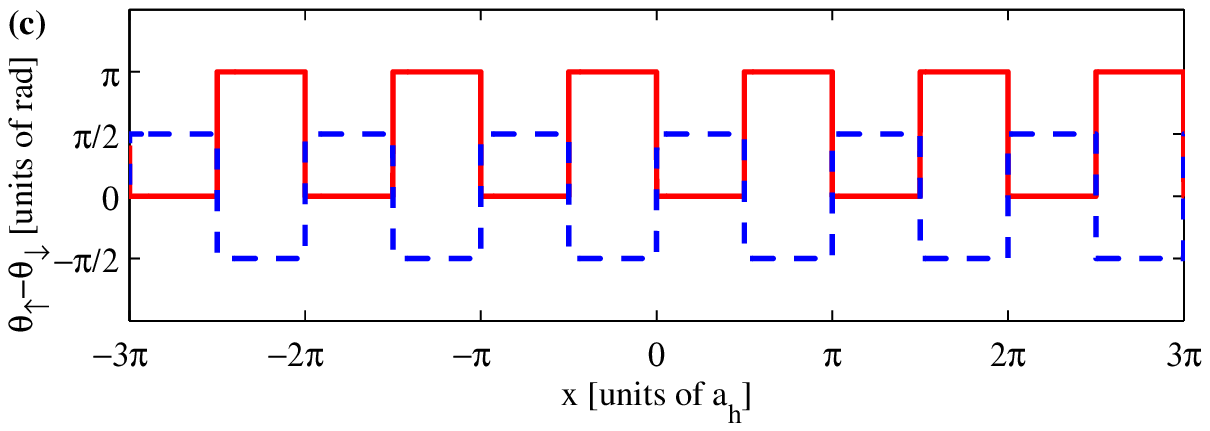} \caption{\label{fig4}(Color online) (a)
The vectorial plots of the pseudospin $\mathbf{S}$ projected
onto the $x$-$y$ plane with $%
\tilde{g}=6000$, $V_{0}=40\hbar\protect\omega_{\perp}$, and
$\kappa=0$ (a1), $\kappa=4a_{h}\protect\omega_{\perp}$ (a2). The
colors ranging from blue to red describe the values of the axial
spin $S_{z}$ from $-1$ to $1$. (b) 3D renderings of N\'{e}el wall
and Bloch wall. (c) Section views of the relative phase
$\theta_{\uparrow}-\theta_{\downarrow}$ along the $x$ axis with
$\kappa=0$ (solid line) and $\kappa=4a_{h}\protect\omega_{\perp}$
(dashed line).}
\end{figure}
\begin{figure}[tbp]
\capstart\centerline{\includegraphics[width=8.5cm,clip=]{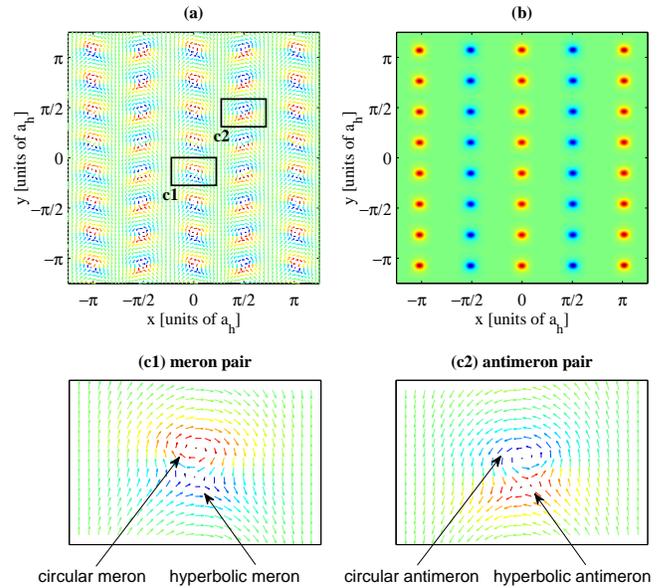}}
\caption{\label{fig5}(Color online) (a) The vectorial plots of the
pseudospin $\mathbf{S}$ projected onto the $x$-$y$ plane under a
pseudo-spin rotation $\sigma _{x}\rightarrow -\sigma _{z}$ and
$\sigma _{z}\rightarrow \sigma _{x}$. (b) The topological charge
density $q(\mathbf{r})$. (c) The amplification of the two kinds of
elements in (a).}
\end{figure}

The alternating arrangement of the vortex and antivortex chains
leads to alternating-direction plane waves, which propagating on two
sides of each chain [See Figs.~\ref{fig3}(b4,b5) and
\ref{fig3}(c4,c5)]. The vortex line density $n_{\text{v}}$ and the
wave number of the plane waves $k_{y}$ satisfy
$n_{\text{v}}=\frac{k_{y}}{\pi}$. Numerical simulations indicate
that for a given SO coupling strength $\kappa $, as the lattice
depth $V_{0}$ increases from 0, $k_{y}$ gradually decreases from
$\frac{m\kappa}{\hbar}$ and eventually becomes $0$ on the boundary
of the vortex lattice phase. This implies that by adjusting the
depth of the optical lattice, one can continuously control the
vortex line density from $0$ to $\frac{m\kappa}{\pi\hbar}$.

\textit{Spin domain wall.---}The separation between the spin-up and
spin-down domains is not sharp, but requires the spin density vector
varying gradually across the opposite domains and forming a spin
domain wall \cite{Malomed}. There are two basic types of domain
walls, N\'{e}el wall and Bloch wall. In the N\'{e}el wall spin flip
occurs in a plane, while in the Bloch wall the spin flip occurs by
tracing a helix [See Figs.~\ref{fig4}(b1) and~\ref{fig4}(b2)]. An
intriguing finding of the present work is that the SO coupling
dramatically changed the domain wall from N\'{e}el wall to Bloch
wall. Figs.~\ref{fig4}(a1) and~\ref{fig4}(a2) show the spin density
vector $\mathbf{S}=\frac{\mathbf{\Psi}
^{\dag}\boldsymbol{\sigma}\mathbf{\Psi}}{|\mathbf{\Psi}|^{2}}$
without and with SO coupling. We can see that in the absence of SO
coupling, the spin density vector across the opposite domains forms
a N\'{e}el wall, while in the presence of SO coupling it forms a
Bloch wall.

This phenomenon can be understood as follows. The direction of the
spin flip in the domain wall only depends on the relative phase, and
can be represented by an azimuthal angle $\alpha =\arctan
(S_{y}/S_{x})=\theta_{\downarrow}-\theta_{\uparrow}$, where $\theta
_{\uparrow}$ and $\theta _{\downarrow}$ are the phases of the wave
functions. When the SO coupling is absent, there is a constant phase
difference $0$ or $\pi $ [See the solid line of Fig.~\ref{fig4}(c)],
so the spin in the walls just flips along the $x$-direction and
forms N\'{e}el walls. When the SO coupling is present, the phase
difference is changed into $\pm \frac{\pi }{2}$ [See the dashed line
of Fig.~\ref{fig4}(c)], so the spin in the walls just flips along
the $y$-direction and forms Bloch walls.

\textit{Meron-pair lattice.---}The regular triangular or rectangular
vortex lattice obtained in Fig.~\ref{fig3} can be equivalently
described by the spin density vector $\mathbf{S}$ in the pseudospin
representation. Fig.~\ref{fig5}(a) presents the vectorial plots of
$\mathbf{S}$ under a pseudo-spin rotation, which corresponding to
the state represented in Figs.~\ref{fig3}(c1-c6), and the
corresponding topological charge
density $q\left( \mathbf{r}\right) =\frac{1}{8\pi }\epsilon ^{ij}\mathbf{S}\cdot \partial _{i}\mathbf{S}%
\times \partial _{j}\mathbf{S}$ is plotted in Fig.~\ref{fig5}(b).
One can see that the spin texture in Fig.~\ref{fig5}(a) represents a
lattice composed of meron pairs and antimeron pairs
\cite{Volovik,Zhou}. Either a meron pair or an antimeron pair has a
``circular-hyperbolic" structure [See Figs.~\ref{fig5}(c1)
and~\ref{fig5}(c2)], and the only difference is that they have
exactly opposite spin orientations. The spatial integral of
$q(\mathbf{r})$ indicates that a meron pair just carries topological
charge $1$, while an antimeron pair carries topological charge $-1$.
Previous studies indicated that stable meron-pair lattice can be
obtained in a rotating system \cite{Kasamatsu2}. Our results show
that meron-pair lattice can also be stabilized in alternating spin
domains by SO coupling without rotation.

\textit{Experimental proposal.---}In real experiments, we can choose
a two-level $^{\text{87}}$Rb BEC system with $\left\vert
F=1,m_{f}=1\right\rangle$\ and $ \left\vert
F=1,m_{f}=-1\right\rangle$. About $N=1.7 \times 10^5$ atoms are
confined in a harmonic trap with the trapping frequencies $\omega
_{\bot }\approx2\pi \times 40$ Hz and $\omega _{z}\approx2\pi \times
200$ Hz. It is convenient to produce spin-dependent optical lattices
with a large lattice spacing by using a CO$_{\text{2}}$ laser
operated at a wavelength of 10.6 $\mu$m. Under typical experimental
conditions, the $s$-wave scattering lengths $a_{ij}\approx100a_{B}$
($a_{B}$ is the Bohr radius). Based on these experimental
parameters,  we can calculate that the effective interaction
parameter $\tilde{g}\approx6000$ and the wave vector
$\nu=a_{h}^{-1}$, which are consistent with our present calculation.

For a given SO coupling strength
$\kappa=4\sqrt{\hbar\omega_{\perp}/m}$, adjusting the lattice depth
from $V_{0}=k_{B}\times 80$ nK  to $V_{0}=k_{B}\times 40$ nK, then
to $V_{0}=k_{B}\times 30$ nK ($k_{B}$ is the Boltzmann's constant),
one can directly observe the phase transitions from density stripe
to triangular vortex lattice, then to rectangle vortex lattice by
monitoring in situ the density profile. And for a given lattice
depth $V_{0}=k_{B}\times 80$ nK, adjusting the SO coupling strength
from $\kappa=0$ to $\kappa=4\sqrt{\hbar\omega_{\perp}/m}$, one may
indirectly observe the transition of the spin domain wall from
N\'{e}el wall to Bloch wall by dual state imaging technique, which
can spatially resolve the relative phase \cite{Anderson}.

\textit{Conclusion.---}In summary, we have investigated the
ground-state phase diagram of spin-orbit coupled BECs in
spin-dependent optical lattices. We observe novel spin-orbit
coupling effects on alternating spin domains produced by the
spin-dependent optical lattices. Actually, alternating spin domains
exist in many physical systems, such as magnetic materials and
condensed matter systems \cite{Vodungbo,Erdin}, thus similar
spin-orbit coupling effects would be discovered in those systems. We
hope that our findings will deepen the understanding of spin-orbit
coupling phenomena and provide thoughts on engineering new quantum
states in spin-orbit coupled systems.

We are grateful to Shih-Chuan Gou for helpful discussions. This work
was supported by the NKBRSFC under Grants No. 2011CB921502, No.
2012CB821305, No. 2009CB930701, and No. 2010CB922904, NSFC under
Grants No. 10972125 and No. 10934010, NSFC-RGC under Grants No.
11061160490 and No. 1386-N-HKU748/10, NSFSP under Grant No.
2010011001-2, and SFRSP.


\begin{thebibliography}{99}
\bibitem{Spielman} Y. J. Lin, K. Jim\'{e}nez-Garc\'{\i}a, and I. B. Spielman, Nature (London) \textbf{471}, 83 (2011).

\bibitem{Jing-Zhang} P. Wang, Z. Q. Yu, Z. Fu, J. Miao, L. Huang, S. Chai, H. Zhai, and J. Zhang, Phys. Rev. Lett. \textbf{109}, 095301 (2012).

\bibitem{Zwierlein} L. W. Cheuk, A. T. Sommer, Z. Hadzibabic, T. Yefsah, W. S. Bakr, and M. W. Zwierlein, Phys. Rev. Lett. \textbf{109}, 095302 (2012).

\bibitem{Shuai-Chen} J. Y. Zhang, S. C. Ji, Z. Chen, L. Zhang, Z. D. Du, B. Yan, G. S. Pan, B. Zhao, Y. J. Deng, H. Zhai, S. Chen, and J. W. Pan, Phys. Rev. Lett. \textbf{109}, 115301 (2012).

\bibitem{Dalibard} J. Dalibard, F. Gerbier, G. Juzeli\={u}nas, and P. \"{O}hberg, Rev. Mod. Phys. \textbf{83}, 1523 (2011).

\bibitem{Niu} A. M. Dudarev, R. B. Diener, I. Carusotto, and Q. Niu, Phys. Rev. Lett. \textbf{92}, 153005 (2004).

\bibitem{Galitski} T. D. Stanescu, B. Anderson, and V. Galitski, Phys. Rev. A \textbf{78}, 023616 (2008).

\bibitem{Wu} X. F. Zhou, J. Zhou, and C. Wu, Phys. Rev. A \textbf{84}, 063624 (2011).

\bibitem{Wu2} Z. Cai, X. Zhou, and C. Wu, Phys. Rev. A \textbf{85}, 061605(R) (2012).

\bibitem{Jung-Hoon-Han} X. Q. Xu and J. H. Han, Phys. Rev. Lett. \textbf{107}, 200401 (2011).

\bibitem{Wuming-Liu} R. Liao, Y. Yi-Xiang, and W. M. Liu, Phys. Rev. Lett. \textbf{108}, 080406 (2012).

\bibitem{Pitaevskii} Y. Li, L. P. Pitaevskii, and S. Stringari, Phys. Rev. Lett. \textbf{108}, 225301 (2012).

\bibitem{Baym} T. Ozawa and G. Baym, Phys. Rev. Lett. \textbf{109}, 025301 (2012).

\bibitem{Trivedi} W. S. Cole, S. Zhang, A. Paramekanti, and N. Trivedi, Phys. Rev. Lett. \textbf{109}, 085302 (2012).

\bibitem{Galitski2} J. Radi\'{c}, A. Di Ciolo, K. Sun, and V. Galitski, Phys. Rev. Lett. \textbf{109}, 085303 (2012).

\bibitem{Wang} C. Wang, C. Gao, C. M. Jian, and H. Zhai, Phys. Rev. Lett. \textbf{105}, 160403 (2010).

\bibitem{Tin-Lun-Ho} T. L. Ho and S. Zhang, Phys. Rev. Lett. \textbf{107}, 150403 (2011).

\bibitem{You} Z. F. Xu, R. L\"{u}, and L. You, Phys. Rev. A \textbf{83}, 053602 (2011).

\bibitem{Santos} S. Sinha, R. Nath, and L. Santos, Phys. Rev. Lett. \textbf{107}, 270401 (2011).

\bibitem{Chuanwei-Zhang} Y. Zhang, L. Mao, and C. Zhang, Phys. Rev. Lett. \textbf{108}, 035302 (2012).

\bibitem{Hu} H. Hu, B. Ramachandhran, H. Pu, and X. J. Liu, Phys. Rev. Lett. \textbf{108}, 010402 (2012).

\bibitem{Ramachandhran} B. Ramachandhran, B. Opanchuk, X. J. Liu, H. Pu, P. D. Drummond, and H. Hu, Phys. Rev. A \textbf{85}, 023606 (2012).

\bibitem{Su-Yi} Y. Deng, J. Cheng, H. Jing, C. P. Sun, and S. Yi, Phys. Rev. Lett. \textbf{108}, 125301 (2012).

\bibitem{Gou} S. W. Su, I. K. Liu, Y. C. Tsai, W. M. Liu, and S. C. Gou, Phys. Rev. A \textbf{86}, 023601 (2012).

\bibitem{Ueda} Z. F. Xu, Y. Kawaguchi, L. You, and M. Ueda, Phys. Rev. A \textbf{86}, 033628 (2012).

\bibitem{Ruokokoski} E. Ruokokoski, J. A. M. Huhtam\"{a}ki, and M. M\"{o}tt\"{o}nen, Phys. Rev. A \textbf{86}, 051607(R) (2012).

\bibitem{Mandel} O. Mandel, M. Greiner, A. Widera, T. Rom, T. W. H\"{a}nsch, and I. Bloch, Phys. Rev. Lett. \textbf{91}, 010407 (2003); Nature (London) \textbf{425}, 937 (2003).

\bibitem{McKay}  D. McKay and B. DeMarco, New J. Phys. \textbf{12}, 055013 (2010).

\bibitem{Becker}  C. Becker, P. Soltan-Panahi, J. Kronj\"{a}ger, S. D\"{o}rscher, K. Bongs and K. Sengstock, New J. Phys. \textbf{12}, 065025 (2010).

\bibitem{Soltan-Panahi}  P. Soltan-Panahi, J. Struck, P. Hauke, A. Bick, W. Plenkers, G. Meineke, C. Becker, P. Windpassinger, M. Lewenstein, and K. Sengstock, Nature Phys. \textbf{7}, 434 (2011).

\bibitem{Soltan-Panahi2}  P. Soltan-Panahi, D. L\"{u}hmann, J. Struck, P. Windpassinger, and K. Sengstock, Nature Phys. \textbf{8}, 71 (2012).

\bibitem{Hauke} P. Hauke, O. Tieleman, A. Celi, C. \"{O}lschl\"{a}ger, J. Simonet, J. Struck, M. Weinberg, P. Windpassinger, K. Sengstock, M. Lewenstein, and A. Eckardt, Phys. Rev. Lett. \textbf{109}, 145301 (2012).

\bibitem{Kasamatsu2} K. Kasamatsu, M. Tsubota, and M. Ueda, Phys. Rev. Lett. \textbf{93}, 250406 (2004).

\bibitem{Kasamatsu} K. Kasamatsu, M. Tsubota, and M. Ueda, Phys. Rev. A \textbf{67}, 033610 (2003).

\bibitem{Malomed} B. A. Malomed, H. E. Nistazakis, D. J. Frantzeskakis, and P. G. Kevrekidis, Phys. Rev. A \textbf{70}, 043616 (2004).

\bibitem{Volovik} G. E. Volovik, \textit{The Universe in a Helium Droplet} (Oxford University, New York, 2003).

\bibitem{Zhou} F. Zhou, Phys. Rev. B \textbf{70}, 125321 (2004).

\bibitem{Anderson} R. P. Anderson, C. Ticknor, A. I. Sidorov, and B. V. Hall, Phys. Rev. A \textbf{80}, 023603 (2009).

\bibitem{Vodungbo} B. Vodungbo, J. Gautier, G. Lambert, A. B. Sardinha, M. Lozano, S. Sebban, M. Ducousso, W. Boutu, K. Li, B. Tudu, M. Tortarolo, R. Hawaldar, R. Delaunay, V. L\'{o}pez-Flores, J. Arabski, C. Boeglin, H. Merdji, P. Zeitoun, and J. L\"{u}ning, Nature Commun. \textbf{3}, 999 (2012).

\bibitem{Erdin} S. Erdin, I. F. Lyuksyutov, V. L. Pokrovsky, and V. M. Vinokur, Phys. Rev. Lett. \textbf{88}, 017001 (2001).

\end{thebibliography}
\end{document}